# Structured light dark-field microscope

SHAOBAI LI,[1] BOFAN SONG,[1] AND RONGGUANG LIANG[1,*]

[1]*James C. Wyant College of Optical Sciences, University of Arizona, 1630 E University Blvd, Tucson, AZ 85721, USA.*
\* *rliang@optics.arizona.edu*

**Abstract:** A resolution-enhanced dark-field microscope by structured light illumination is proposed to improve resolution and contrast. A set of phase-shifted fringes are projected to the sample plane at large angle to capture modulated dark-field images, from which resolution- and contrast-enhanced dark-field image, as well as sectioned dark-field image, can be obtained. Human tissue samples are tested to demonstrate the resolution and contrast enhancement. The system can be implemented in transmission-mode and reflectance-mode, with potential applications ranging from defect detection to biomedical imaging.

## 1. Introduction

Dark-field microscopy is a simple and effective technique for biological samples imaging, especially for rendering unstained and transparent specimens [1–3]. It is a commonly used optical method to obtain edge-enhanced images of objects with gradients in either amplitude or phase. The traditional method is to illuminate the object at an angle larger than the acceptance angle of the objective. Various dark-field condensers have been developed and commercialized for transillumination and reflectance imaging modes. With a separated dark-field condenser, a regular microscope can be used for dark-field imaging. In reflectance imaging mode, some special objectives with integrated dark-field illumination features have been developed and commercialized as well.

In general, the dark-field image lacks the low spatial frequencies of the sample, rendering the image a high-passed version of the underlying structure. The resolution of the dark-field image is determined by the numerical aperture of the objective. If the system is not well-designed and built, the image contrast will be degraded due to the stray light. Since spatial resolution is key parameter for microscopy, several techniques have been proposed to improve resolution or contrast for dark-field imaging. Jiang et al. proposed a nanoscale solid immersion lenses (n-SILs) based strategy to achieve super resolution under dark-field illumination [4]. Olshausen et al combined evanescent illumination via total internal reflection (TIR) with dark-field microscopy to improve lateral resolution [5]. However, those techniques are complex, difficult to implement, and not suitable for general applications.

In this paper, we propose a resolution-enhanced dark-field microscope by using structured light illumination. Structured illumination microscopy (SIM) is a widely used imaging technique to improve the image resolution and contrast. It was first introduced to eliminate the out-of-focus background encountered in wide-field microscopy and to improve its signal-to-noise ratio [6]. By projecting a set of phase-shifted sinusoidal fringe patterns onto an object and the high-contrast sectioned image can be computed from the fringe-modulated images of the object. A new super-resolution method based on structured light illumination was developed to improve the lateral resolution by a factor of two in its linear form and by a larger factor in a nonlinear form [7-9]. The improvement of resolution is obtained by shifting high frequency components of the object spectrum into the microscope's optical transfer function domain through a frequency mixing process.

## 2. Structured light dark-field microscope in transmission mode

The proposed system can work in both transmission-mode and reflectance-mode for dark-field imaging. The transmission-mode dark-field microscope is shown in Fig. 1. The structured light pattern from the digital micromirror device (DMD) projector (DLP® Discovery™ 4100 Development Platform) is first imaged by lens L1(Thorlabs AC508-150-A-ML, focal length=150 mm) to an intermediated image, which is then collimated by lens L2 (Doublet, focal length=500 mm). A ring aperture is used to block the central part of the light, only the outer region of the light can pass through and be focused onto the sample plane by the objective O1 (NA=0.75) to create dark-field illumination. With this illumination configuration, the DMD array is imaged onto the object at a large angle. Therefore, a set of phase-shifted fringe patterns can be projected onto the object. The objective O2 with a smaller aperture (NA=0.4) than that of the objective O1 collects the scattering light from the sample and blocks the direct illumination light. A darkfield image is formed on the detector (IDS, UI-3080SE-C-HQ, 2456 x 2054-pixel resolution, 3.45 μm pixel pitch) plane by Lens L3 (Thorlabs AC254-150-A-ML, focal length=150 mm).

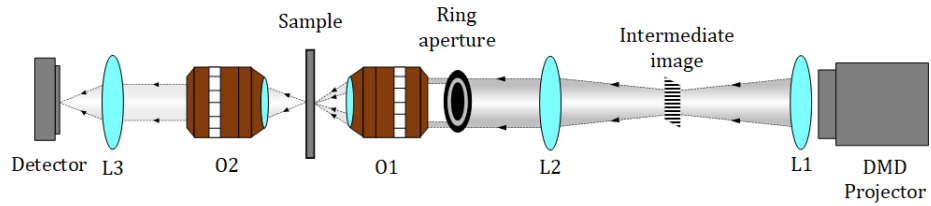

Fig. 1. Experiment setup of the transmission-mode structured light dark-field microscope. L1-L3: lens; O1-O2: objective lens.

In a conventional dark-field microscope, the spatial frequency bandwidth that can be detected by an objective is described by the optical transfer function (OTF), which supports the highest spatial frequencies up to a specific cutoff frequency. Structured illumination techniques deliver better resolution than conventional microscope by encoding extra high spatial frequencies structural details into the sample, and reconstruct the high-resolution image from a set of low-frequency images via spatial frequency mixing [7]. To test the performance of the system, two test targets are imaged. 12 fringe patterns with 4 orientations ($\theta = 0°$, $45°$, $90°$, $135°$) and 3 phase shifts ($\varphi = 0°$, $120°$, $240°$) are generated by the DMD projector and projected onto the sample, the fringe frequency is 300 cycles/mm in the DMD plane. Figures 2(a) and 2(b) show the conventional and resolution-enhanced dark-field images of the bars in Group 7 of the 1951 USAF resolution target. The reconstructed dark-field image in Fig. 2(b) shows sharper edges and better contrast than the conventional dark-field image in Fig. 2(a). Figures 2(c) and 2(d) are the conventional and reconstructed resolution-enhanced dark-field images of the FocalCheck ring in the microscope test slide (Thermo Fisher Scientific Inc, F36909 FocalCheck™ fluorescence microscope test slide #1). The thin ring in Fig. 2(c) is blurred and the contrast is low; however, the resolution-enhanced dark-field image in Fig. 2(d) has a much sharper edge and a much better contrast. Figures 2(e) and 2(f) plot the cross-section profiles of the marked red lines in Figs. 2(b) and 2(d), showing the significant resolution enhancement of the proposed dark-field imaging technique.

We also demonstrate the system performance with unstained human tissue samples (US Biolab Corporation Inc, Rockville, MD 20850). Figures 3(a1), 3(b1), and 3(c1) are the bright field images of brain, skin, and oral tissue samples. It is very clear that bright field images have the lowest image resolution and contrast, this is the reason that advanced imaging techniques, such as phase contrast imaging and dark-field imaging, are needed. Another solution is to stain the tissue, which is a time-consuming process. Dark-field imaging method is sometimes used to examine the unstained tissue samples because the dark-field microscope is simple, easy to use, and low-cost. Figures 3(a2), 3(b2), and 3(c2) are the corresponding conventional dark-field images, showing the better image resolution and contrast, compared to the bright field images. Figures 3(a3), 3(b3), and 3(c3) are the corresponding resolution-enhanced dark-field images. The image resolution and contrast are significantly improved, as evidenced by the highlighted regions in Figs. 3(a2) and 3(a3) as the examples.

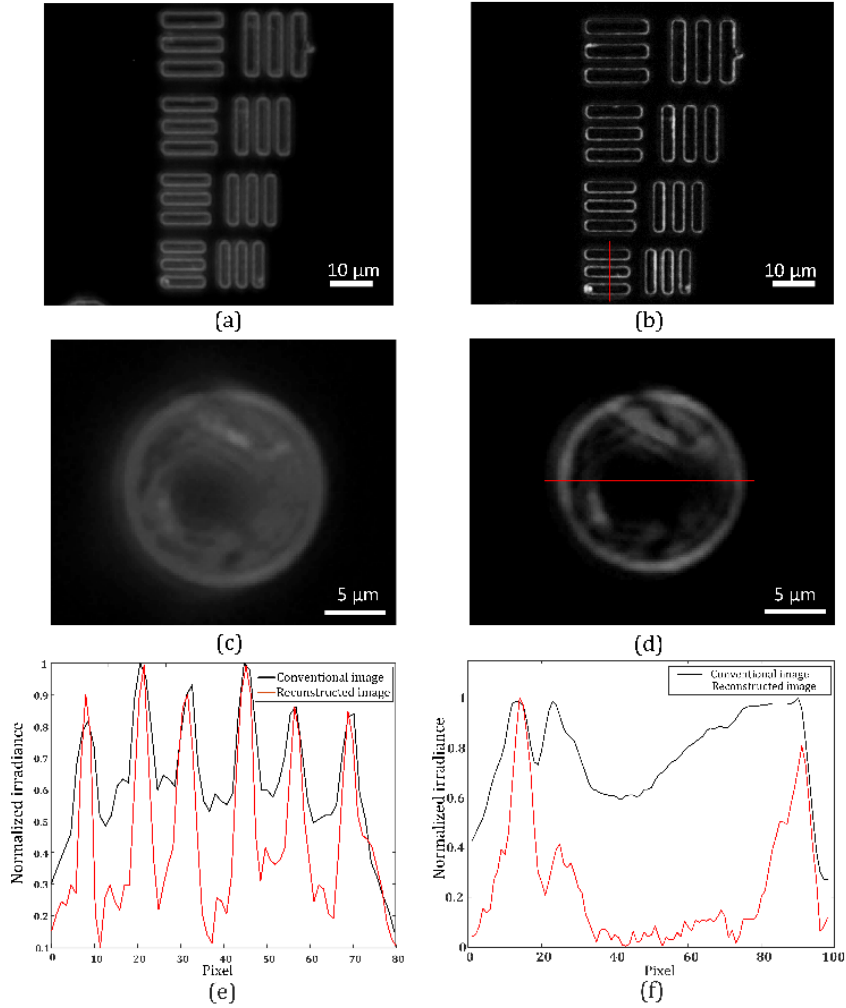

Fig. 2. Experimental validation of the structured light dark-field microscope. (a) and (b) are the conventional and the reconstructed dark-field images of Group 7 pattern in 1951 USAF resolution target, (c) and (d) are the conventional and reconstructed dark-field images of a thin ring in the microscope test slide, (e) and (f) are the corresponding cross-section profiles of the marked red lines in (b) and (d).

With 3 phase shifted fringe modulated dark-field images, we can also obtain sectioned dark-field images, as shown in Figs. 3(a4), 3(b4), and 3(c4). The sectioned image is calculated as $\sqrt{(I_1 - I_2)^2 + (I_2 - I_3)^2 + (I_3 - I_1)^2}$ , where $I_1, I_2,$ and $I_3$ are the intensity images modulated by three phase-shifted sinusoidal fringe patterns [6]. The overall image resolution and contrast of sectioned images are better than the conventional dark-field images, but not as good as the resolution-enhanced dark-field images because only the information in the focal plane are enhanced.

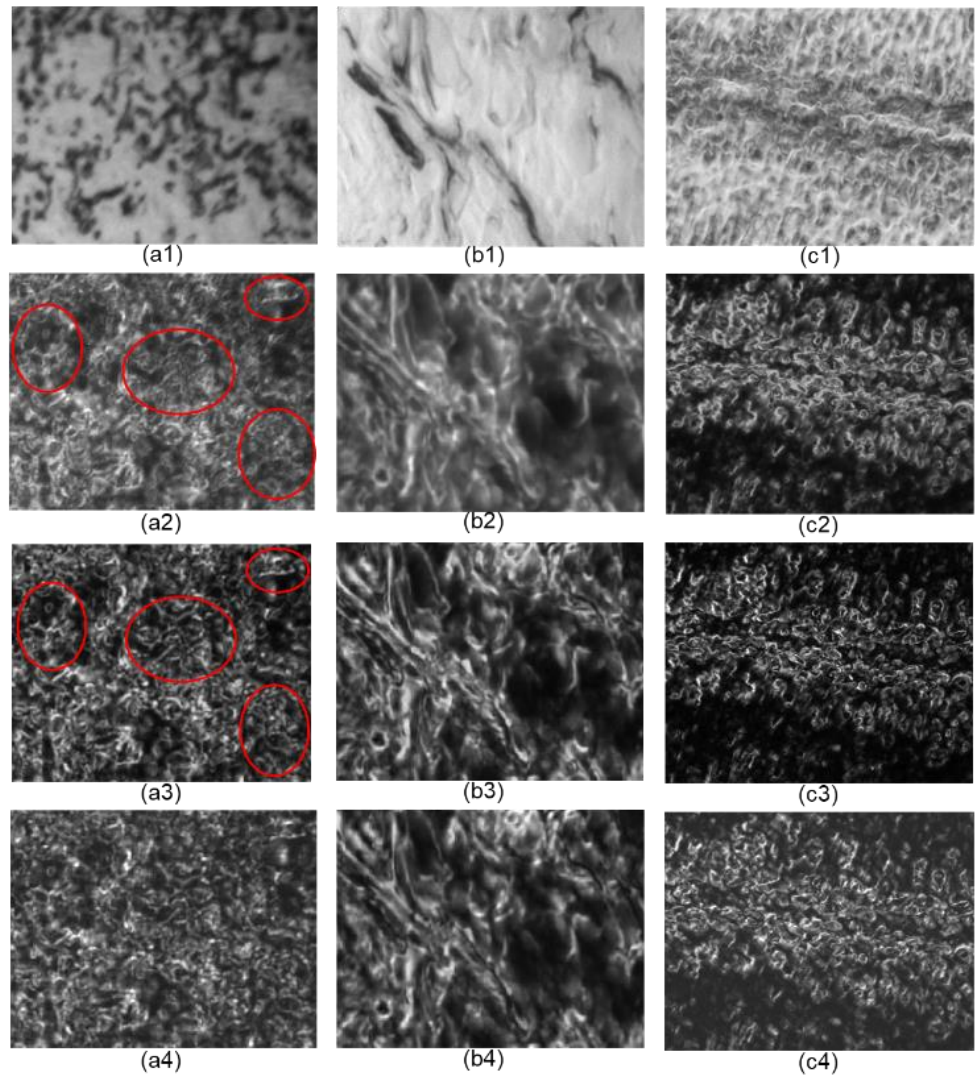

Fig. 3. Experimental validation of the structured light dark-field microscope. (a1), (b1) and (c1) are the conventional bright-field images of brain, skin, and oral tissue samples. (a2), (b2), and (c2) are the corresponding conventional dark-field images, showing the better image resolution and contrast. (a3), (b3), and (c3) are the corresponding resolution-enhanced dark-field images. (a4), (b4), and (c4) are the corresponding sectioned dark-field images.

## 3. Structured light dark-field microscope in reflectance mode

The structured illumination dark-field imaging can also be implemented in reflectance-mode as shown in Fig. 4. The structured light from the projector is first imaged by lens L1 to an intermediate image and then focused to the sample after passing through lenses L2, L3, L3, and objective O1. A 4-f system (L3 and L4) images the ring aperture to the aperture plane of the objective lens O1. The scattering light from the sample is then collected by the microscope objective O1 again and is imaged to the detector. The second 4-f system (L4 and L5) images the aperture plane of the objective to an intermediate plane where an aperture is placed to block the light directly reflected from the sample.

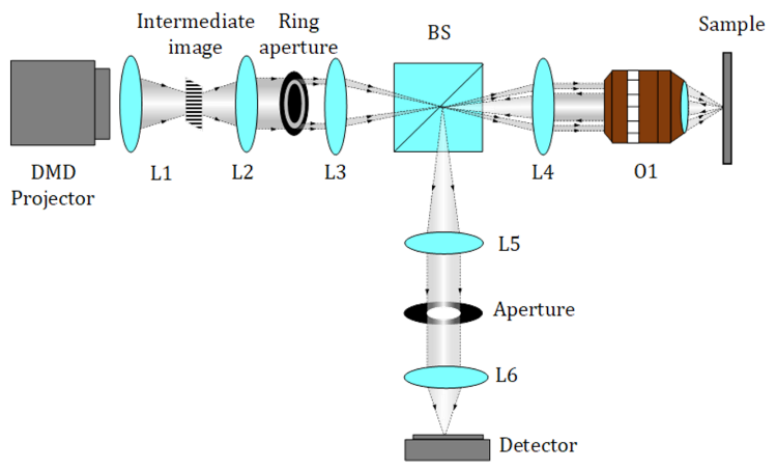

Fig. 4. Experiment setup of the reflectance-mode structured light dark-field microscope. L1-L6: lens; BS: beam splitter; O1: Objective lens.

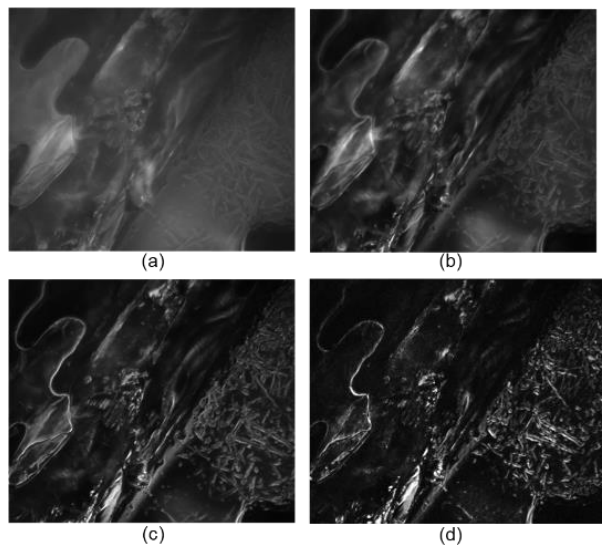

Fig. 5. Experimental results of the structured-light dark-field imaging in reflectance-mode. (a), (b), (c), and (d) are the bright field, conventional dark-field, resolution-enhanced dark-field, and sectioned dark-field image of the Potassium Ferricyanide slide.

Figures 5(a), 5(b), 5(c), and 5(d) are the reflectance bright field, conventional dark-field, resolution-enhanced dark-field, and sectioned dark-field images of the Potassium Ferricyanide slide in the reflectance-mode. As expected, the conventional dark-field image is able to reveal some details that the reflectance bright field image fails to provide. The resolution-enhanced dark-field has a much better resolution and contrast than the conventional dark-field image. The sectioned dark-field image highlights

some features in the focused plane, but the overall resolution and contrast are not as good as the resolution-enhanced dark-field image.

## 4. Conclusion

In this paper, a resolution-enhanced dark-field microscope by using structured light illumination is introduced and demonstrated. The proposed system can be implemented in both transmission and reflectance modes. The performance of the proposed method has been validated with 1951 USAF resolution target and demonstrated with human tissue samples. The resolution and contrast of the dark-field image can be significantly improved by using structured light illumination. The system is easy to implement and will have great potential applications from industry to biomedical fields.

## Disclosures

The authors declare that there are no conflicts of interest related to this article.

**References**

1. Horio, Tetsuya, and Hirokazu Hotani. "Visualization of the dynamic instability of individual microtubules by dark-field microscopy," *Nature* 321(6070), 605-607 (1986).
2. R. M. Macnab, "Examination of bacterial flagellation by dark-field microscopy," *Journal of clinical microbiology*, *4*(3), pp.258-265 (1976).
3. Fangyao Hu, Robert Morhard, Helen A. Murphy, Caigang Zhu, and Nimmi Ramanujam, "Dark field optical imaging reveals vascular changes in an inducible hamster cheek pouch model during carcinogenesis," Biomed. Opt. Express 7, 3247-3261 (2016).
4. Liyong Jiang, Wei Zhang, Hua Yuan, and Xiangyin Li, "Super resolution from pure/hybrid nanoscale solid immersion lenses under dark-field illumination," Opt. Express 24, 25224-25232 (2016).
5. Philipp von Olshausen and Alexander Rohrbach, "Coherent total internal reflection dark-field microscopy: label-free imaging beyond the diffraction limit," Opt. Lett. 38, 4066-4069 (2013).
6. M. A. A. Neil, R. Juškaitis, and T. Wilson, "Method of obtaining optical sectioning by using structured light in a conventional microscope," Opt. Lett. 22, 1905-1907 (1997).
7. M. G. Gustafsson, "Surpassing the lateral resolution limit by a factor of two using structured illumination microscopy," *Journal of microscopy*, *198*(2), 82-87 (2000).
8. Dan, Dan, Ming Lei, Baoli Yao, Wen Wang, Martin Winterhalder, Andreas Zumbusch, Yujiao Qi et al. "DMD-based LED-illumination super-resolution and optical sectioning microscopy." *Scientific reports* 3, no. 1, 1-7 (2013).
9. Vicente Micó, Juanjuan Zheng, Javier Garcia, Zeev Zalevsky, and Peng Gao, "Resolution enhancement in quantitative phase microscopy," Adv. Opt. Photon. 11, 135-214 (2019).